\documentclass[12pt]{iopart}

\usepackage[cp1251]{inputenc}
\usepackage{iopams}
\usepackage{graphicx,epsfig}
\begin{document}

\title[Wave function, Einstein's boxes and scattering a particle on a 1D $\delta$-potential]{New about the wave function,
``Einstein's boxes'' and scattering a particle on a one-dimensional $\delta$-potential}

\author{N. L. Chuprikov}

\address{Tomsk State Pedagogical University, 634041, Tomsk, Russia}
\ead{chnl@tspu.edu.ru} \vspace{10pt}

\begin{abstract}

The connection between the problem of scattering a particle on a one-dimensional $\delta$-potential with the ``Einstein's
boxes'' thought experiment is shown. In both cases, the validity of the superposition principle is limited by Einstein's
'separation principle'. It is also shown that the generally accepted point of view, according to which ``To know the
quantum mechanical state of a system implies, in general, only statistical restrictions on the results of measurements'',
is fundamentally wrong. First, even the square of the modulus of the wave function imposes more than just statistical
restrictions. Second, the phase of the wave function also has a physical meaning -- it sets the field of pulses of the
ensemble. That is, quantum mechanics not only does not prohibit the simultaneous measurement of the coordinate and
momentum of a particle, but also predicts the value of the momentum at that spatial point where the particle will
(accidentally) be detected.

\end{abstract}


\newcommand{\ppp}{\mbox{\hspace{5mm}}}
\newcommand{\ppa}{\mbox{\hspace{15mm}}}
\newcommand{\ppb}{\mbox{\hspace{20mm}}}
\newcommand{\ooo}{\mbox{\hspace{3mm}}}
\newcommand{\ooa}{\mbox{\hspace{1mm}}}

\section{Instead of Introduction} \label{int}
\subsection{Tunneling time problem and ``Einstein's Boxes''} \label{tunbox}

\hspace*{\parindent} At present, in the question of describing the scattering of a quantum particle on a one-dimensional
$\delta$-potential barrier, a rather contradictory situation has developed. On the one hand, the modern quantum mechanical
model (SQM) of this process \cite{Alb} has been developed at a rigorous mathematical level, in full compliance with the
principles of quantum mechanics and, it would seem, should provide an exhaustive explanation of all its physical
properties. On the other hand, this process remains a mystery for researchers (see, for example, \cite{Ste0,Ste}), since
all attempts to investigate its temporal aspects on the basis of SQM have not been crowned with success. The so-called
tunneling time problem (TTP), which arises for any one-dimensional short-range potential barrier, turned out to be so
confused that not only is there still no generally accepted definition of the tunneling time, but even the reason why it
cannot be solved remains unclear.

Note that the SQM itself has never been questioned. But, as will be shown below, the reason lies precisely in it, since
this model is contradictory (and this applies to the model for any other one-dimensional short-range potential barrier).
To reveal the essence of this contradiction, we begin with one of the key provisions of the SQM, which describes the
process with one-sided incidence of a particle onto a barrier, according to which all scattering states in this problem
are asymptotically free as $t\to\mp\infty$. This means that in these two limiting cases the particle does not interact
with the barrier, and the region of its localization consists of two disjoint spatial intervals separated by a barrier
region in which the probability of finding the particle is strictly zero. For example, in the case of the
$\delta$-potential, ``given'' at the point $x=0$, two disjoint intervals are the intervals $(-\infty,0)$ and $(0,\infty)$,
separated by a point $x=0$.

But the same situation arises in ``Einstein's Boxes'' thought experiment, which Einstein used against the orthodox
doctrine of the completeness of quantum mechanics (see \cite{Fine,Bal2,How,Nor,Held,Bri} and also \cite{Miel0,Miel}). In
the scattering problem, division into two ``boxes'' occurs in the scattering process at $t\to\mp\infty$, and their role is
played by the intervals $(-\infty,0)$ and $(0,\infty)$. In this regard, the search for the cause will begin with an
analysis of the standard quantum-mechanical description of this experiment (although in its original version it was about
a ball, we will consider the version with a particle).

\subsection{``Einstein's boxes'': separation principle versus superposition principle} \label{twobox}

According to Einstein, the standard (orthodox) quantum mechanical description of this experiment contains a contradiction,
the essence of which he formulated as follows (quoted from \cite{How,Nor}): {\it ``the paradox forces us to relinquish one
of the following two assertions:
\begin{itemize}
\item[(1)] the description by means of the $\psi$-function is complete,
\item [(2)] the real states of spatially separated objects are independent of each other.
\end{itemize}
On the other hand, it is possible to adhere to (2) if one regards the $\psi$-function as the description of a
(statistical) ensemble of systems (and therefore relinquishes (1)). However, this view blasts the framework of the
'orthodox quantum theory'.''}

But, in our opinion, this formulation inaccurately reflects the essence of the contradiction that arises when describing
this experiment within the orthodox approach (following the \cite{How}, statement (2) we will consider as the ``separation
principle''). First, assertion (2) contains a {\it sufficient}, but not a {\it necessary} condition for incompleteness of
quantum mechanics: assertion (1) really contradicts assertion (2); but $\psi$-function gives an incomplete description of
the particle dynamics in the case of a single box too (it would be strange if in an experiment with two boxes the wave
function gave an incomplete description, and in an experiment with one box it did a complete one). And here it should be
recalled (see \cite{Held}) that Einstein himself did not always use the separation principle to prove the incompleteness
of quantum mechanics.

Second, this formulation is based on the assumption (see also \cite{Bal}) that the transition from the orthodox
interpretation of the wave function to the statistical interpretation (SI) is quite sufficient to reconcile quantum
mechanics with the separation principle. However, it is not. What actually contradicts in the traditional
quantum-mechanical description of an experiment with two boxes isolated from each other is its provision that the
superposition of two pure states of a particle localized in the independent boxes should also be considered as a pure
state. And what is important, this contradicts not only position (2), which Einstein formulated, but also the definition
of mixed states, which is inherent in quantum mechanics itself.

Let's dwell on this issue in more detail. Let $\cal{H}$ be the Hilbert space associated with this thought experiment. And
let $\psi_1$ and $\psi_2$ be two unit vectors from $\cal{H}$, which describe the states of the particle localized in the
first and second boxes, respectively. Consider the superposition
\begin{eqnarray} \label{1000}
\fl \psi=\alpha_1 e^{i\lambda_1}\psi_1+\alpha_2 e^{i\lambda_2}\psi_2,
\end{eqnarray}
where $\lambda_1$ and $\lambda_2$ are arbitrary (real) phases, and real parameters $\alpha_1$ and $\alpha_2$ satisfy the
condition $\alpha_1^2+\alpha_2^2=1$. The question arises: ``Is this a pure state or a mixed state?''

According to ortodox quantum mechanics, all state vectors from $\cal{H}$ and their superposition describe pure states, and
any observable $O$ corresponds to a self-adjoint operator $\hat{O}$, whose domain of definition is everywhere dense in
$\cal{H}$. Therefore, on the one hand, the $\psi$ state should be treated as pure, and the average
$\langle\psi|\hat{O}|\psi\rangle$ should depend on all four parameters, including the $\lambda_1$ phases and $\lambda_2$.

On the other hand, if we take into account Exp. (\ref{1000}) for $\psi$ and the fact that the states $\psi_1$ and $\psi_2$
are localized in two disjoint spatial regions (the formulation of a thought experiment assumes that they are separated
infinitely deep potential well) and, thus, $\langle\psi_1|\hat{O}|\psi_2\rangle=0$, then
\begin{eqnarray}\label{1001}
\fl \langle\psi|\hat{O}|\psi\rangle=\alpha_1^2\cdot \langle\psi_1|\hat{O}|\psi_1\rangle+\alpha_2^2\cdot
\langle\psi_2|\hat{O}|\psi_2\rangle.
\end{eqnarray}
But, according to quantum mechanics, such expansions are typical of mixed states, so the superposition (\ref{1000}) cannot
be considered a pure state (the fact that $\langle\psi|\hat{O}|\psi\rangle$ does not depend on the phases $\lambda_1$ and
$\lambda_2$ is also one of the hallmarks of a mixed state). Thus, in order to reconcile the quantum-mechanical description
of this thought experiment with the definition of mixed states and with the separation principle, it is necessary to
revise the existing position that all vectors from $\cal{H}$ describe pure states; in other words, it is necessary to
revise the existing formulation of the principle of superposition in relation to the ``Einstein's boxes''.

\subsection{The transition from a pure state into a mixed state in a closed system} \label{onetwo}

However, having settled the contradiction that appears in the standard description of the state of a particle in two
boxes, we arrive at something else. The fact is that in this experiment the particle is first located in one box, and only
then this box, with the help of an absolutely impenetrable partition, is divided into two boxes independent of each other.
This means that a mixed state in this experiment arises only at the final stage, and at the initial moment of time, when
there is only one box, the particle is in a pure state, since Einstein assumed that the state of the particle in the
experiment is described by the $\psi$-function. Thus, in the course of this experiment, the initially pure state of the
particle is transformed into a mixed one. But the fact is that the transition from a pure state to a mixed one, according
to orthodox quantum mechanics, is possible only in open systems, thanks to the decoherence process, which is described
using a reduced density matrix.

At first glance, in the case of Einstein's Boxes, this contradiction is easily resolved. First, although Einstein assumed
that the state of a particle is described by a $\psi$-function, an exact quantum-mechanical analysis of the dynamics of a
particle, as a closed system, is impossible within the framework of his purely speculative  setting of this problem.
Secondly, dividing the original box into two using an ideal impenetrable partition involves the intervention of an
external agent. But this means that in this experiment the particle cannot be considered a closed system; the process of
decoherence occurs precisely due to this external interference.

Thus, the thought experiment itself cannot be viewed as a serious challenge for the traditional scenario of the system's
transition from a pure state to a mixed state (in fact, this is another reason why Einstein's criticism was not properly
received). However, such a scenario does not work in the case of a particle scattering on a one-dimensional $\delta$
-potential barrier -- an analogue of the Einstein Boxes. At all stages of this process, the particle is a closed system,
the state of which is described by the wave function. Thus, the SQM -- the existing model of this process -- must be
revised, since the appearance of a mixed state when $t\to\infty$  should inevitably lead, in the Schrodinger formalism, to
a restriction of the superposition principle in the corresponding Hilbert state space and, consequently, to new physics of
this process.

Note that the concept of a mixed vector state is itself unusual in Schrodinger's formalism. Therefore, before to proceed
to a revision of the SQM, we must first understand what restrictions the wave function imposes on the physical properties
of this unusual quantum state and how these restrictions differ from those imposed on a pure state and a mixed state
determined with the density operator. In doing so, we must proceed from the doctrine of the incompleteness of quantum
mechanics, according to which quantum mechanics is the theory of quantum statistical ensembles. In this regard, we will
adhere to the statistical interpretation (SI) \cite{Bal}, which, in our opinion, most accurately reflects the essence of
quantum mechanics.

In principle, we could resolve this question during the revision of the SQM. However, we are forced to highlight it
separately, since at present quantum mechanics gives a false idea of these limitations even in the case of an ordinary,
pure quantum state. For example, in the very first phrase of his book \cite{Bell}, John Bell writes {\it ``To know the
quantum mechanical state of a system implies, in general, only statistical restrictions on the results of measurements''}.
This idea, based on the Born interpretation of the wave function, is so ingrained in modern quantum mechanics that it is
considered an immutable fact. It is this idea that underlies long debate on the foundations of quantum mechanics.

Against this background, the Bohm approach stands out. In the works \cite{Bohm1,Bohm2} David Bohm showed that if we
introduce into quantum mechanics an additional postulate on the existence of single-particle trajectories, then the phase
of the wave function determines the particle momentum, and its modulus determines not only statistical restrictions but
also the quantum-mechanical potential. As a consequence, in this approach, the quantum mechanical state implies not only
{\it statistical} restrictions on the measurement results.

It is generally accepted that the wave function acquires these properties only due to the postulate about trajectories.
And since this postulate does not fit into the framework of quantum mechanics, Bohm's finds were not taken, anyhow, into
account in other interpretations of quantum mechanics, including SR. At the same time, as will be shown in the next
section, these finds, with corresponding corrections, also arise in SI, where there is no postulate about trajectories.

\section{Physical meaning of the modulus and phase of the wave function} \label{ensemble}
\subsection{Pure state} \label{pure}

Let us begin with a pure state
\begin{eqnarray}\label{1100}
\fl \psi(x,t)=\sqrt{w(x,t)}\ooa e^{i\phi(x,t)},
\end{eqnarray}
where $w(x,t)$ and $\phi(x,t)$ are real functions; $\int_{-\infty}^\infty w(x,t) dx=1$; $w(x,t)\geq 0$. According to Max
Born, $w(x,t)=|\psi(x,t)|^2$ is the probability density, and in SI (see \cite{Bal}) it is a function describing the
distribution of particles along the coordinate $x$ at time $t$ in the corresponding quantum ensemble. As for the phase
$\phi(x,t)$, nothing is said about its physical meaning in SI. To find it out, we write the average value of the momentum
operator $\hat{p}=-i\hbar\frac{d}{dx}$ in the form
\begin{eqnarray}\label{1002}
\fl \langle p\rangle=\int_{-\infty}^\infty \psi^*(x,t)\hat{p}\psi(x,t) dx\equiv\int_{-\infty}^\infty p(x,t) w(x,t) dx,
\end{eqnarray}
where
\begin{eqnarray}\label{1003}
\fl p(x,t)=\frac{\mathrm{Re}\left(\psi^*(x,t)\hat{p}\psi(x,t)\right)}{w(x,t)}=\frac{\hbar}{w(x,t)}
\mathrm{Im}\left(\psi^*(x,t)\psi_x(x,t)\right)=\hbar\phi_x(x,t)
\end{eqnarray}
for any function $f(x,t)$, hereinafter, $f_x(x,t)\equiv\partial f(x,t)/\partial x$ and $f_{xx}(x,t)\equiv\partial^2
f(x,t)/\partial x^2$. Thus, not only the function $w(x,t)$ -- the square of the modulus of the wave function $\psi(x,t)$
-- which we will call below the 'probability field', has a physical meaning, but also its phase $\phi(x,t)$, which defines
the field of pulses $p(x,t)$; both fields characterize the physical properties of the quantum ensemble.

Note that the definition of the impulse variable in terms of the phase of the wave function, similar to the definition
(\ref{1003}), also appears in Bohm quantum mechanics \cite{Bohm1}. But in \cite{Bohm1} the coordinate $x$ depends on $t$,
and the momentum is defined as the particle momentum on a {\it trajectory} $x(t)$. In our approach, the function $p(x,t)$
depends on two {\it independent} variables $x$ and $t$, and it is introduced on the basis of a standard formula in quantum
mechanics for calculating the average momentum over an ensemble of particles (see the first equality in Section
(\ref{1002})).

For a quantum ensemble, in addition to the momentum field $p(x,t)$, the kinetic energy field $K(x,t)$ can be introduced.
For this, we write the average value of the kinetic energy operator $\hat{K}=\hat{p}^2/2m$ in the form
\begin{eqnarray}\label{1004}
\fl \langle K\rangle=\frac{1}{2m}\int_{-\infty}^\infty \psi^*(x,t)\hat{p}^2\psi(x,t) dx\equiv\int_{-\infty}^\infty K(x,t)
w(x,t) dx,
\end{eqnarray}
where
\begin{eqnarray}\label{1005}
\fl K(x,t)=\frac{1}{2m}\left[p(x,t)\right]^2-\frac{\hbar^2}{4m}\left[\frac{w_{xx}(x,t)}{w(x,t)}-
\frac{1}{2}\left(\frac{w_{x}(x,t)}{w(x,t)}\right)^2\right].
\end{eqnarray}
As is seen, the function $K(x,t)$ contains two contributions: in a sense, the first contribution has a 'corpuscular'
nature, and the second is of a 'wave' nature. It is interesting to note that the expression for the second contribution is
the same as for the Bohm 'quantum mechanical potential'. But in our approach, this contribution has a different physical
meaning. Common to both approaches is the fact that the modulus of the wave function is related not only to probability.

Obviously, in the case of a spinless particle, for a quantum ensemble, one can introduce a field of any physical quantity
$O$ with a self-adjoint operator $\hat{O}$:
\begin{eqnarray*}\label{1006}
\fl \langle O\rangle=\int_{-\infty}^\infty \psi^*(x,t)\hat{O}\psi(x,t) dx\equiv\int_{-\infty}^\infty O(x,t) w(x,t) dx;
\ooo O(x,t)=\frac{\mathrm{Re}\left[\psi^*(x,t)\hat{O}\psi(x,t)\right]}{\psi^*(x,t)\psi(x,t)}
\end{eqnarray*}
Therefore, the information inherent in the wave function about the physical properties of 'pure' quantum ensembles is much
richer than it has been assumed until now.

Note, in order to experimentally verify the predictions of quantum mechanics regarding the properties of a pure quantum
ensemble, there is no need to test all the fields that characterize it. For this, it is sufficient to investigate only the
probability field $w(x,t)$, which is related to the modulus of the wave function, and the momentum field $p(x,t)$, which
is related to its phase. Similarly, in the momentum representation we have the probability field $w(p,t)$, the coordinate
field $x(p,t)$ and so on.

A special situation arises in those problems when the stationary states of the particle corresponding to the eigenvalues
of the energy operator from the discrete spectrum are real. In this case, the phase of the wave function is zero. As a
consequence, the field of pulses is also equal to zero, and the field of kinetic energy is determined only by the 'wave'
term in (\ref{1005}).

\subsection{Mixed state defined by superposition of pure states}

Now consider a mixed state (\ref{1000}), which is a superposition of two pure states. Let the wave function $\psi_1(x,t)$
define a (pure) quantum ensemble with a nonzero distribution function $w_1(x,t)$ in the spatial domain $G_1$ and a
momentum field $p_1(x,t)$. Accordingly, let the wave function $\psi_2(x,t)$ define a (pure) quantum ensemble with a
nonzero distribution function $w_2(x,t)$ in the spatial domain $G_2$ disjoint with the domain $G_1$, and field impulses
$p_2(x,t)$.

Since both ensembles are localized in non-overlapping spatial regions, according to the separation principle, the state
(\ref{1000}) is mixed. This means that experimental verification of the predictions of quantum mechanics for this state is
reduced to independent testing of the fields of pure states $\psi_1(x,t)$ and $\psi_2(x,t)$. It makes no physical sense to
calculate the average values of physical quantities for the superposition of these two states.

\subsection{Mixed state specified by the density operator}

For comparison, now consider the mixed state, which is specified by the density operator
$\hat{\rho}=c_1|\psi_1\rangle\langle\psi_1| + c_2|\psi_2\rangle\langle\psi_2|$, where $c_1,\ooa c_2\geq 0$, $c_1 + c_2
=1$. In contrast to the previous case, we will assume that the pure states $|\psi_1\rangle$ and $|\psi_2\rangle$ are now
localized in the same spatial region, but the designations for the probability field and the momentum field of each state
will remain the same.

Now, when testing a quantum ensemble in state $\hat{\rho}$ at time $t$, with probability $c_1\cdot w_1(x,t)dx$, we will
find a particle in the interval $[x,x + dx]$ with momentum $p_1(x,t)$, and with probability $c_2\cdot w_2(x,t)dx$ we will
find it in the same interval with momentum $p_2(x,t)$. That is, in the case of a mixed ensemble given by the density
operator $\hat{\rho}$, when measuring the momentum of a particle accidentally detected in the interval $[x,x + dx]$, now
we can obtain (with probabilities $c_1$ and $c_2$) two values, and not one, as in the case of a pure state (\ref{1100})
and a mixed state (\ref{1000}) (in this case, the interval $[x,x+dx]$ can only belong to one of the regions $G_1$ and
$G_2$).

\section{Scattering a particle on a one-dimensional $\delta$-potential barrier in the context of ``Einstein's boxes''} \label{start}

Now let us study in more detail the connection between particle scattering on a one-dimensional $\delta$-potential barrier
with the thought experiment ``Einstein's Boxes'' and present a new model of this process.

\subsection{Stationary scattering states} \label{start}

Let us consider the $\delta$-potential $V(x)=W\delta(x)$ where $W>0$ (there are no bound states). According to the SQM,
the stationary Schr\"{o}dinger equation can be written as
\begin{eqnarray} \label{1}
\fl \hat{H}_{tot}\Psi_{tot}(x,k)\equiv
-\frac{\hbar^2}{2m}\frac{d^2\Psi_{tot}(x,k)}{dx^2}+W\delta(x)\Psi_{tot}(x,k)=E\Psi_{tot}(x,k);
\end{eqnarray}
where $k=\sqrt{2mE}/\hbar$ and $E$ is the particle energy; the Hamiltonian $\hat{H}_{tot}$ corresponds to the self-adjoint extension $H_{\kappa,0}$ of the densely defined
symmetrical operator $\dot{H}=-\frac{\hbar^2}{2m}\frac{d^2}{dx^2}$ with $\mathrm{Dom}(\dot{H})=\{g\in H^{2,2}(\mathbb{R})|g(0)=0\}$ (see p.75 in \cite{Alb}). The corresponding
boundary conditions are
\begin{eqnarray} \label{2}
\fl \Psi_{tot}(0^+)=\Psi_{tot}(0^-),\ppp \Psi_{tot}^\prime (0^+)-\Psi_{tot}^\prime(0^-)=2\kappa\Psi_{tot}(0^-);
\end{eqnarray}
hereinafter, $\kappa=mW/\hbar^2$ and $f(0^\pm)=\lim_{\epsilon\to 0}f(\pm \epsilon)$ for any function $f(x)$; the prime
denotes a derivative.

There are other two self-adjoint extensions of the operator $\dot{H}$ in the SQM. However, they are considered there as
special cases corresponding to $\kappa=0$ and $\kappa=\infty$ and, unlike $H_{\kappa,0}$, play the secondary role in this
model. An intrigue is that in our approach we face with the opposite situation. We show that, in fact, these two special
cases have no relation to $\kappa=0$ and $\kappa=\infty$ and, unlike $H_{\kappa,0}$, play a key role (see Section
\ref{part}) in the description of this scattering process.

The eigenvalues of the operator $\hat{H}_{tot}$ are doubly degenerate and lie in the domain $E\geq 0$. Thus, the general
solution to the equation (\ref{1}) with the boundary conditions (\ref{2}) can be written as a linear superposition of two
linearly independent particular solutions. As such, functions
\begin{eqnarray} \label{3}
\fl \Psi_{tot}^{L}(x,k)=\left\{
\begin{array}{rl}
e^{ikx}+A_{ref}(k)e^{-ikx};\ooo x<0\\
\fl A_{tr}(k)e^{ikx};\ooo x>0
\end{array} \right.\\
\fl \Psi_{tot}^{R}(x,k)=\left\{
\begin{array}{rl}
A_{tr}(k)e^{-ikx};\ooo x<0\\
e^{-ikx}+A_{ref}(k)e^{ikx};\ooo x>0
\end{array} \right.\nonumber
\end{eqnarray}
are usually taken, which describe a particle incident on the barrier from the left and right, respectively; here $A_{tr}(k)=k/(k+i\kappa)$, $A_{ref}(k)=-i\kappa/(k+i\kappa)$.
The quantities $T(k)=|A_{tr}(k)|^2=k^2/(k^2+\kappa^2)$ and $R(k)=|A_{ref}(k)|^2=\kappa^2/(k^2+\kappa^2)$ represent the transmission and reflection coefficients, respectively
(note that the transfer matrix for the delta potential can be obtained as the limiting transfer matrix of a rectangular potential barrier if the width of the barrier tends to
zero and its area is fixed). As is seen, $T(0)=0$. Therefore the functions $\Psi_{tot}^{L}(x,k)$ and $\Psi_{tot}^{R}(x,k)$ are identically zero for $k=0$. Thus, the ground
states are not involved in the construction of (non-stationary) scattering states -- there are no particles with zero momentum in the quantum (one-particle) ensemble of
particles incident on the barrier.

\subsection{Scattering states with asymptotically free dynamics} \label{nonstart}

Our next step is to find time-dependent solutions to the Schr\"{o}dinger equation which would describe free dynamics in
the limiting cases $t\to\mp\infty$. For only such states can be considered ``scattering states''. For a particle incident
on the barrier from the left, such states can be written in the form
\begin{eqnarray} \label{14L}
\fl \Psi_{tot}^L(x,t)=\frac{1}{\sqrt{2\pi}}\int_{-\infty}^\infty \mathcal{A}(k,t) \Psi_{tot}^L(x,k) dk;
\end{eqnarray}
where $\mathcal{A}(k,t)=\mathcal{A}_{in}(k) \exp[i(ka-E(k)t/\hbar)]$; a real function $\mathcal{A}_{in}(k)$ is such that
the norm of the left asymptote
\begin{eqnarray} \label{14f}
\fl \Psi^L_{in}(x,t)=\frac{1}{\sqrt {2\pi}}\int_{-\infty}^\infty\mathcal{A}(k,t)e^{ikx}dk
\end{eqnarray}
is equal to one: $\int_{-\infty}^\infty [\mathcal{A}_{in}(k)]^2 dk=1$. At the initial instant of time $t=0$, the peak of
the wave packet $\Psi^L_{in}(x,t)$ is located at the point $x=-a$. Accordingly, for a particle incident on the barrier
from the right, a time-dependent scattering state is
\begin{eqnarray} \label{14R}
\fl \Psi_{tot}^R(x,t)=\frac{1}{\sqrt{2\pi}}\int_{-\infty}^\infty \mathcal{A}(k,t) \Psi_{tot}^R(x,k) dk.
\end{eqnarray}
In this case, the norm of the right in-asymptote
\begin{eqnarray} \label{14r}
\fl \Psi^R_{in}(x,t)=\frac{1}{\sqrt{2\pi}} \int_{-\infty}^\infty\mathcal{A}(k,t)e^{-ikx}dk
\end{eqnarray}
is equal to one, and its peak is at the point $x=+a$.

Let us consider, as $\mathcal{A}_{in}(k)$, the Gaussian function $\mathcal{A}_{in}(k)\equiv\mathcal{A}_G(k)=c\ooa
e^{-L^2(k-k_0)^2}$, where $c=\sqrt[4]{\frac{2L^2}{\pi}}$, $L$ is the width of the wave packet, in the $k$-space this wave
packet is peaked at the point $k_0$. Strictly speaking, such a choice of the function $\mathcal{A}_{in}(k)$ does not meet
the important requirement of the physical formulation of the scattering problem, since the wave packets (\ref{14f}) and
(\ref{14r}) must be constructed only from waves that move towards the barrier. This means that $\mathcal{A}_{in}(k)$ must
be nonzero only for $k>0$. As will be shown below, the function $\mathcal{A}_G(k)$ satisfies this condition only in some
limiting cases.

According to the SQM (see \cite{Alb,Re3}), there is a strong limit in this scattering problem, and therefore the norms of
$\Psi_{tot}^L(x,t)$ and $\Psi_{tot}^R(x,t)$, like the norms of their in-asymptotes, must be equal to one. Let us check
this property by the example of the state (\ref{14L}) under the assumption that $\mathcal{A}_{in}(k)$ is nonzero and for
$k\leq 0$:
\begin{eqnarray}\label{14b}
\fl \langle\Psi_{tot}^L|\Psi_{tot}^L\rangle=\frac{1}{2\pi}\int_{-\infty}^\infty  \int_{-\infty}^\infty
[\mathcal{A}(k',t)]^* \mathcal{A}(k,t) I(k',k) dk' dk;\nonumber\\
\fl I(k',k)=\lim_{X\to\infty} \tilde{I}(X,k',k);\ppp \tilde{I}(X,k',k)=\int_{-X}^X [\Psi_{tot}^L(x,k')]^*\Psi_{tot}^L(x,k)
dx.
\end{eqnarray}
Substituting Exp. (\ref{3}) for $\Psi_{tot}^L(x,t)$ in (\ref{14b}), we get
\begin{eqnarray*}
\fl \tilde{I}(X,k',k)=\frac{2(k'k+\kappa^2)+i\kappa(k'-k)}{(k'-i\kappa)(k+i\kappa)}\ooa\frac{\sin[(k'-k)X]}{k'-k}
-\frac{i\kappa (k'-k-2i\kappa)}{(k'-i\kappa)(k+i\kappa)}\ooa\frac{\sin[(k'+k)X]}{k'+k}\\
\fl +\frac{2\kappa}{(k'-i\kappa)(k+i\kappa)}
\left[\sin^2\left(X\frac{k'+k}{2}\right)-\sin^2\left(X\frac{k'-k}{2}\right)\right].
\end{eqnarray*}

Further, given that $\lim_{X\to\infty} \frac{\sin(kX)}{k}=\pi\delta(k)$ and $x\delta(x)=0$, we get
\begin{eqnarray*}
\fl I(k',k)=\frac{2(k'k+\kappa^2)+i\kappa(k'-k)}{(k'-i\kappa)(k+i\kappa)}\ooa\pi \delta(k'-k)
-\frac{i\kappa(k'-k-2i\kappa)}{(k'-i\kappa)(k+i\kappa)}\ooa \pi\delta(k'+k)\\
\fl =2\pi\delta(k'-k)-\frac{2\pi i\kappa}{k+i\kappa}\delta(k'+k).
\end{eqnarray*}
Thus,
\begin{eqnarray} \label{19}
\fl \langle\Psi_{tot}^L|\Psi_{tot}^L\rangle=\int_{-\infty}^\infty [\mathcal{A}_{in}(k)]^2 dk-\int_{-\infty}^\infty
\mathcal{A}_{in}(-k)\mathcal{A}_{in}(k)e^{2ika}\frac{i\kappa}{k+i\kappa}dk\nonumber\\
\fl =\langle\Psi_{in}^L|\Psi_{in}^L\rangle+\kappa\int_{-\infty}^\infty \mathcal{A}_{in}(-k) \mathcal{A}_{in}(k)
\frac{k\sin(2ka)-\kappa \cos(2ka)}{k^2+\kappa^2} dk;
\end{eqnarray}
here we took into account that $\mathcal{A}_{in}(-k)\mathcal{A}_{in}(k)$ is an even real function. A similar situation
arises in the case of the state (\ref{14R}).

Thus, $\langle\Psi_{tot}^L|\Psi_{tot}^L\rangle=\langle\Psi_{in}^L|\Psi_{in}^L\rangle$ when $\mathcal{A}_{in}(-k)\mathcal{A}_{in}(k)\equiv 0$. This takes place when
$\mathcal{A}_{in}(k)\in C_0^\infty(\mathbb{R}\backslash\{0\})=C_0^\infty(-\infty,0)\oplus C_0^\infty (0,\infty)$, where $C_0^\infty (-\infty,0)$ and $C_0^\infty(0,\infty)$ are
the subspaces of infinitely differentiable functions which are identically zero on the semi-axises $[0,\infty)$ and $(-\infty,0]$, respectively; for $|k|\to 0$ they tend to zero
faster than $|k|^n$; for $|k|\to \infty$ they tend to zero faster than $1/|k|^n$; $n$ is a positive integer. With such functions $\mathcal{A}_{in}(k)$, solutions
$\Psi_{tot}^L(x,t)$ and $\Psi_{tot}^R(x,t)$ of the time-dependent Schr\"{o}dinger equation describe the scattering states with asymptotically free dynamics.

Another situation arises when $\mathcal{A}_{in}(k)$ is the Gaussian function $\mathcal{A}_G(k)$. Now (\ref{19}) can be
rewritten in the form
\begin{eqnarray} \label{20}
\fl \langle\Psi_{tot}^L|\Psi_{tot}^L\rangle=\langle\Psi_{in}^L|\Psi_{in}^L\rangle-\sqrt{2\pi}\kappa L \ooa\mbox{erfc}\left(\frac{2\kappa
L^2+a}{\sqrt{2}L}\right)e^{2L^2(\kappa^2-k_0^2)+2\kappa a}.
\end{eqnarray}
As is seen, the interference term is approximately zero when $a/L\gg 1$, $L\kappa\gg 1$, $L k_0\gg 1$. But if we also take
into account that, at the initial instant of time, particles in the quantum ensemble must move towards the barrier, then
the restrictions on the parameters of the Gaussian function $\mathcal{A}_{in}(k)$ will be written in the form $a\gg L\gg
1/k_0$. That is, the wave packets $\Psi^L_{in}(x,t)$ and $\Psi^R_{in}(x,t)$ should be quasi-monochromatic, and the width
of each of these packets at $t=0$ should be much less than the distance between the packet peak and the barrier.

\subsection{A scattering particle with asymptotically free dynamics as an analogue of a particle
in ``Einstein boxes''} \label{nonst}

According to the SQM, each scattering state has one in-asymptote and one out-asymptote, and these asymptotes are not
related to other scattering states. This property is satisfied when the Hamiltonian $\hat{H}_{tot}$ is indeed self-adjoint
(the corresponding quantum dynamics is unitary and (hence) unique). However, in this scattering problem, asymptotically
free scattering states do not possess this property.

Consider the state $\Psi_{tot}^L(x,t)$ with the Gaussian function $\mathcal{A}_{in}(k)$ for which the interference term in
(\ref{20}) is negligible. The advantage of making use of such states compared to scattering states with functions
$\mathcal{A}_{in}(k)$ from the space $C_0^\infty(\mathbb{R}\backslash \{0\})$ is that in this case the wave function
$\Psi_{tot}^L(x,t)$ can be found in analytical form.

So, let $\mathcal{A}_{in}(k)=\mathcal{A}_G(k)$ in (\ref{14L}). Then, taking into account (\ref{3}), we obtain
\begin{eqnarray} \label{33}
\fl \Psi_{tot}^{L}(x,t)=\left\{
\begin{array}{rl}
\Psi_{in}^L(x,t)-i\kappa G(-x,t);\ooo x<0\\
\fl \Psi_{in}^L(x,t)-i\kappa G(x,t);\ooo x>0
\end{array} \right.
\end{eqnarray}
where $\Psi_{in}^L(x,t)$ is the in-asymptote (see (\ref{14f}))
\begin{eqnarray} \label{53}
\fl \Psi_{in}^L(x,t)=\frac{c}{\sqrt{2(L^2+ibt)}} \exp\left(\frac{-(x+a)^2+4ik_0 L^2(x+a-b k_0 t)}{4(L^2+ibt)}\right),
\end{eqnarray}
$b=\hbar/(2m)$; and
\begin{eqnarray} \label{52}
\fl G(x,t)=\frac{1}{\sqrt{2\pi}}\int_{-\infty}^\infty \mathcal{A}(k,t) \frac{e^{ik(x+a)}}{k+i\kappa}dk.
\end{eqnarray}
The integral $G(x,t)$ can be found as a solution to the equation
\begin{eqnarray*}
\fl \frac{\partial G(x,t)}{\partial x}=\kappa G(x,t)+i \Psi_{in}^L(x,t)
\end{eqnarray*}
which follows from (\ref{52}). It can be shown that
\begin{eqnarray}\label{54}
\fl G(x,t)=-ic\sqrt{\frac{\pi}{2}}\ooa \mbox{erfc}\left(\frac{x+a-2 iL^2 k_0}{2\sqrt{L^2+ibt}}+\kappa \sqrt{L^2+ibt}\right)e^{L^2(\kappa-i k_0)^2+i b \kappa^2 t+\kappa(x+a)}.
\end{eqnarray}
For what follows, we also need the integral
\begin{eqnarray} \label{52a}
\fl F(x,t)=\frac{1}{\sqrt{2\pi}}\int_{-\infty}^\infty \mathcal{A}(k,t) \frac{e^{ik(x+a)}}{k-i\kappa}dk.
\end{eqnarray}
It is easy to show that
\begin{eqnarray*}
\fl F(x,t)=ic\sqrt{\frac{\pi}{2}}\ooa \mbox{erfc}\left(-\frac{x+a-2 iL^2 k_0}{2\sqrt{L^2+ibt}}+\kappa \sqrt{L^2+ibt}\right)e^{L^2(\kappa+i k_0)^2+i b \kappa^2 t-\kappa(x+a)}.
\end{eqnarray*}

Now we have, in analytical form, not only the scattering state (\ref{33}) itself and its in-asymptote (\ref{53}), but also
its out-asymptote which represents a superposition
\begin{eqnarray} \label{55a}
\fl \Psi_{out}(x,t)= \Psi_{out}^{L}(x,t)+ \Psi_{out}^{R}(x,t)
\end{eqnarray}
of the left and right asymptotes $\Psi_{out}^{L}(x,t)$ and $\Psi_{out}^{R}(x,t)$,
\begin{eqnarray} \label{55}
\fl \Psi_{out}^{L}(x,t)= -i\kappa G(-x,t),\ppp \Psi_{out}^{R}(x,t)= \Psi_{in}^L(x,t)-i\kappa G(x,t),
\end{eqnarray}
localized in the non-intersecting spatial regions lying on the opposite sides of the barrier.

According to the SQM, only the scattering state (\ref{33}) is related to this asymptote. But this is not the case. Let us consider the family of the stationary states
\begin{eqnarray}  \label{63}
\Psi(x,k;\lambda)=\Psi_{tot}^{L}(x,k)+(e^{i\lambda}-1)\tilde{\Psi}(x,k)
\end{eqnarray}
with different values of the parameter $\lambda$, where
\begin{eqnarray*} \label{56a}
\fl \tilde{\Psi}(x,k)=\left\{
\begin{array}{rl}
\frac{k^2}{k^2+\kappa^2} e^{ikx};\ooo x<0\\
\frac{k}{k+i\kappa} e^{ikx}+\frac{ik\kappa}{k^2+\kappa^2} e^{-ikx};\ooo x>0
\end{array} \right.
\end{eqnarray*}
The corresponding scattering states built with the Gaussian function $\mathcal{A}_G(k)$ are
\begin{eqnarray} \label{56b}
\Psi(x,t;\lambda)=\Psi_{tot}^{L}(x,t)+(e^{i\lambda}-1)\tilde{\Psi}(x,t),
\end{eqnarray}
where
\begin{eqnarray*} \label{56}
\fl \tilde{\Psi}(x,t)=\left\{
\begin{array}{rl}
\Psi_{in}^L(x,t)-\frac{i\kappa}{2}[G(x,t)-F(x,t)];\ooo x<0\\
\Psi_{in}^L(x,t)-i\kappa G(x,t)+\frac{i\kappa}{2}[G(-x,t)+F(-x,t)];\ooo x>0
\end{array} \right.
\end{eqnarray*}
Their out-asymptotes (coinciding at $\lambda=0$ with the out-asymptote (\ref{55a})) are
\begin{eqnarray} \label{57}
\fl \Psi_{out}(x,t;\lambda)= \Psi_{out}^{L}(x,t)+ e^{i\lambda}\Psi_{out}^{R}(x,t),
\end{eqnarray}
localized, in the limit $t\to\infty$, in the disjoint spatial regions lying on the opposite sides of the barrier. That is,
this situation is similar to the one that occurs in ``Einstein's boxes''.

\begin{figure}[h]
\begin{center}
\includegraphics[width=7.0cm,angle=0]{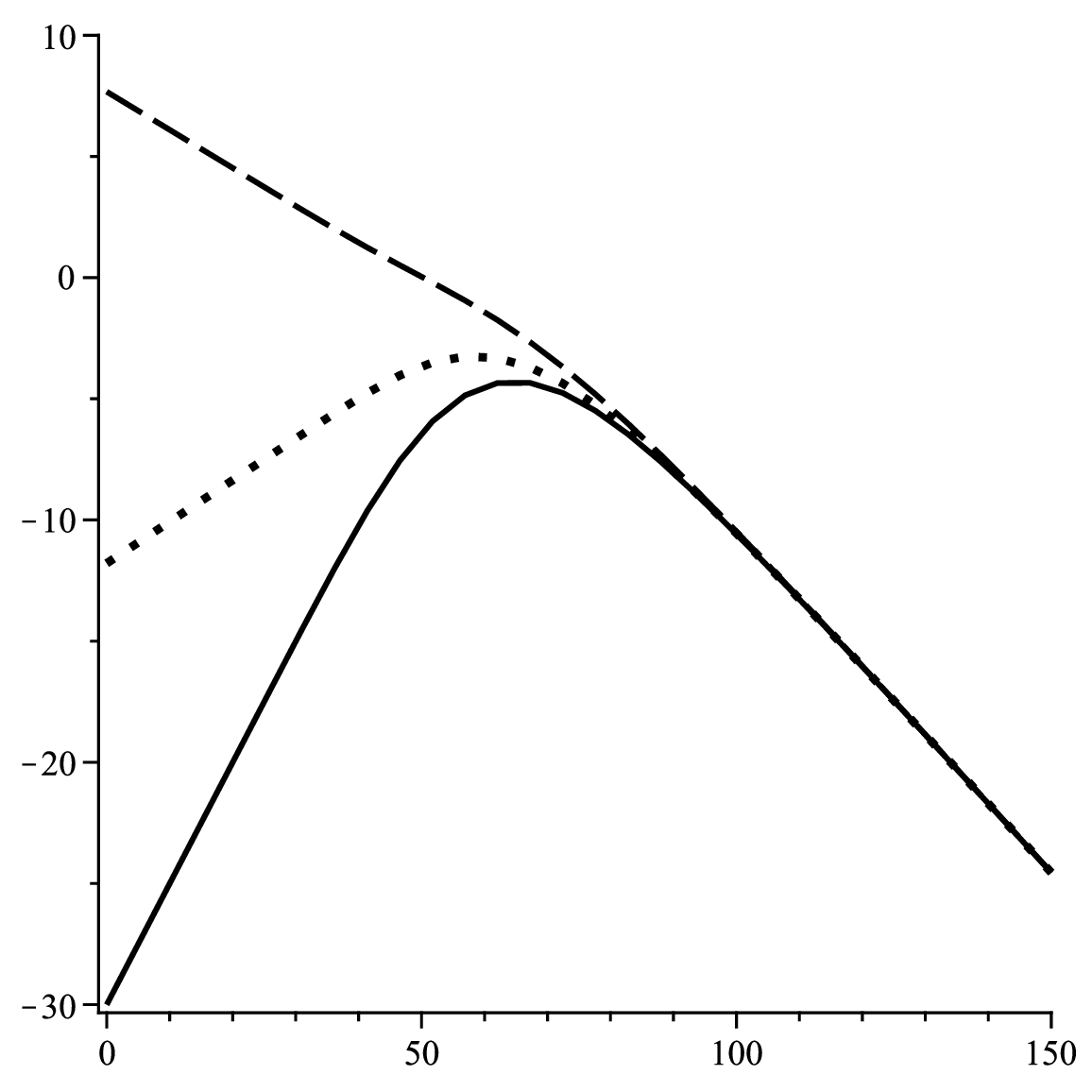}
\end{center}
\caption{Dependence of $\langle x\rangle$ on $t$ for $L=5$, $a=30$, $k_0=0.5$, $\kappa=1$: $\lambda=0$ -- solid line,
$\lambda=\pi/2$ -- dotted line, $\lambda=\pi$ -- dashed line; $\hbar=m=1$.} \label{fig:fig1}
\end{figure}

Due to the barrier region that separates the out-asymptotes $\Psi_{out}^{L}(x,t)$ and $\Psi_{out}^{R}(x,t)$ in the limit
$t\to\infty$, the mean value of any observable does not depend on the phase $\lambda$ (of course, this property is
strictly fulfilled only for $\mathcal{A}_{in}(k)\in C_0^\infty(\mathbb{R}\backslash\{0\})$). This is demonstrated on
fig.~\ref{fig:fig1} by the example of the average value of the position operator, calculated for the state
$\Psi_{out}(x,t;\lambda)$ at $\lambda=0$, $\lambda=\pi/2$ and $\lambda=\pi$; making use of the units $\hbar=m=1$, we take
$\kappa=1$, $L=5$, $a=30$ and $k_0=0.5$. It should be stressed that the interference term in (\ref{20}) is negligible in
this case.

In the limiting case $t\to \infty$, for the out-asymptote (\ref{57}) it is valid the averaging rule
\begin{eqnarray*}
\fl \langle \Psi_{out}|x|\Psi_{out}\rangle\equiv\langle x\rangle_{out}=\langle T\rangle\cdot \langle
x\rangle_{out}^R+\langle R\rangle\cdot \langle x\rangle_{out}^L
\end{eqnarray*}
which says that $\lambda$ is unobserved and the state (\ref{57}) is mixed; here
\begin{eqnarray*}
\fl \langle x\rangle_{out}^R=\frac{\langle \Psi_{out}^R|x|\Psi_{out}^R\rangle}{\langle \Psi_{out}^R|\Psi_{out}^R\rangle},
\ooa \langle x\rangle_{out}^L=\frac{\langle \Psi_{out}^L|x|\Psi_{out}^L\rangle}{\langle \Psi_{out}^L|\Psi_{out}^L\rangle};
\ooo T=\langle \Psi_{out}^R|\Psi_{out}^R\rangle,\ooa R=\langle \Psi_{out}^L|\Psi_{out}^L\rangle;
\end{eqnarray*}
$T$ and $R$ are the transmission and reflection coefficients.

So, for a particle that falls on the barrier from the left and is described by a wave function (\ref{33}), in the limit
$t\to \infty$, exactly the same situation arises as for a particle in the experiment with two boxes. The 'boxes' here are
the intervals $(-\infty, 0)$ and $(0, \infty)$; the out-asymptote $\Psi_{out}^{L}(x,t)$, which describes reflected
particles, is localized in the left 'box', and the out-asymptote $\Psi_{out}^{R}(x,t)$, which describes the transmitted
particles, is localized in the right 'box'. Thus, the superposition of these two (pure) out-asymptotes, localized in
independent 'boxes', is a mixed state (analogous to the state (\ref{1000})). And since this 'mixed' out-asymptote is
common for the entire one-parameter class of states $\Psi(x,t;\ lambda)$ given by the expression (\ref{56b}), we come to
the conclusion that all non-stationary states (\ref{56b}) are mixed states at $t\to \infty$.

Note that the states $\Psi(x,t;\lambda)$ (\ref{56b}) at $\lambda \neq 0$ describe scattering processes with two-sided
incidence of a particle on a barrier. In this case, the analogy with two boxes arises not only in the limit $t\to\infty$,
but also in the limit $t\to -\infty$. Thus, the states $\Psi(x,t;\lambda)$ for $\lambda\neq 0$ are mixed ones also for
$t\to \infty$.

As for the state $\Psi(x,t;0) = \Psi_{tot}^L(x,t)$, which describes the scattering problem with a left-hand fall of a
particle onto a barrier, its in-asymptote $\Psi_{in}^L(x,t)$ is a pure state. Thus, a situation arises that is paradoxical
for the Schrodinger formalism -- the nonstationary scattering state $\Psi_{tot}^L(x,t)$ interpolates between the pure
in-asymptote $\Psi_{in}^L(x,t)$ and the superposition of two independent out-asymptotes $\Psi_{out}^{L}(x,t)$ and
$\Psi_{out}^{R}(x,t)$, which is a mixed state.

Such Schrodinger dynamics could be viewed as a transition of a closed system from a pure state to a mixed one (see section
\ref{onetwo}). But, as shown above, this dynamics ceases to be unambiguous when $t\to \infty$. Therefore, it is not
unitary, which means that the operator $\hat{H}_{tot}$ is not self-adjoint, and no observable can be defined for the
entire scattering process. In other words, the unsteady state of scattering $\Psi(x,t;0)$ (that is, $\Psi_{tot}^L(x,t)$)
is a mixed state, and the process itself with a left-hand fall of a particle onto a barrier is is a mixture of two
sub-processes -- transmission and reflection. In this regard, it becomes necessary to reconstruct the entire prehistory of
the transmitted and reflected wave packets to describe the subprocesses at all stages of scattering.

As shown in \cite{Chup} using a rectangular potential barrier as an example, for any one-dimensional short-range potential
barrier this prehistory can be reconstructed uniquely by the in-asymptote $\Psi_{in}^L(x,t)$, which describes the entire
ensemble of particles, and out-asymptotes $\Psi_{out}^{L}(x,t)$ and $\Psi_{out}^{R}(x,t)$ that describing sub-ensembles of
transmitted and reflected particles (this idea is also applicable to the potential step). In the case of the
$\delta$-potential, a modification of this approach is possible, which we present in the next two sections.

\subsection{Self-adjoint extensions associated with the 'periodic' and Dirichlet boundary conditions} \label{part}

Let us consider the presented in \cite{Alb} two 'special cases' of self-adjoint extensions of the operator $\dot{H}$. One
of them involves the periodic boundary conditions
\begin{eqnarray} \label{9a}
\fl \psi(0^+)=\psi(0^-),\ppp \psi^\prime(0^+)=\psi^\prime(0^-).
\end{eqnarray}
The corresponding (self-adjoint) Hamiltonian $\hat{H}_0$ (see \cite{Alb}) will be also denoted by $\hat{H}_{tr}$:
\begin{eqnarray} \label{9b}
\fl \hat{H}_{tr}=\hat{H}_0= -\frac{d^2}{dx^2};\ppp Dom(\hat{H}_0)=W^2_2(\mathbb{R}).
\end{eqnarray}
Note that $\hat{H}_0\neq\lim_{\kappa\to 0}H_{\kappa, 0}$ because $Dom(H_{\kappa, 0})\neq Dom(\hat{H}_0)$ at any arbitrary smal value of $\kappa>0$. For a free particle, two
independent solutions of the corresponding stationary Schr\"{o}dinger equation are
\begin{eqnarray} \label{11}
\fl \Psi_{tr}^{L}(x,k)=e^{ikx},\ppp \Psi_{tr}^{R}(x,k)=e^{-ikx};\ppp x\in(-\infty,\infty).
\end{eqnarray}

Another 'special case' is associated with the Dirichlet boundary conditions
\begin{eqnarray} \label{9}
\fl \psi(0^+)=\psi(0^-)=0.
\end{eqnarray}
The corresponding self-adjoint extension of $\dot{H}$ will be denoted by $\hat{H}_{ref}$. Note (see also \cite{Alb}), the boundary conditions (\ref{9}) do not impose any
restrictions on the derivatives $\psi^\prime(0^+)$ and $\psi^\prime(0^-)$, thereby totally disconnecting the $x$-intervals $(-\infty,0)$ and $(0,\infty)$. Thus,
$\hat{H}_{ref}\neq \lim_{\kappa\to \infty} H_{\kappa,0}$ because the boundary conditions (\ref{2}) do not disconnect these intervals even in the limit $\kappa\to\infty$. So,
\begin{eqnarray} \label{10a}
\fl \hat{H}_{ref}=\hat{H}^L_{ref}\oplus\hat{H}^R_{ref},
\end{eqnarray}
and the eigenfunctions of the operators $\hat{H}^L_{ref}$ and $\hat{H}^R_{ref}$ are defined on the semi-axes $(-\infty, 0)$ and $(0,\infty)$, respectively. Solutions to the
corresponding stationary Schr\"{o}dinger equations are
\begin{eqnarray} \label{10}
\fl \Psi_{ref}^{L}(x,k)= e^{ikx}-e^{-ikx},\ooo x<0;\ppp \Psi_{ref}^{R}(x,k)=e^{-ikx}-e^{ikx},\ooo x>0.
\end{eqnarray}

\subsection{Scattering states as coherent superpositions of transmission and reflection states} \label{alt}

Let us now show that the state $\Psi_{tot}^L(x,k)$ can be uniquely represented as a superposition of the states $\Psi_{tr}^L(x,k)$ and $\Psi_{ref}^L(x,k)$. For this purpose, let
us write the incident wave of the state $\Psi_{tot}^{L}(x,k)$ as a superposition of two incident waves, with the amplitudes $A^{tr}_{in}(k)$ and $A^{ref}_{in}(k)$, associated
with the states $\Psi_{tr}^{L}(x,k)$ and $\Psi_{ref}^{L}(x,k)$, respectively. In this case, we will assume that $A^{tr}_{in}(k)=|A_{tr}(k)|e^{i\mu(k)}$ and
$A^{ref}_{in}(k)=|A_{ref}(k)|e^{i\nu(k)}$. Real phases $\mu$ and $\nu $ obey the equation $\sqrt{T(k)}e^{i\mu(k)}+\sqrt{R(k)} e^{i\nu(k)}=1$ which has two roots
\begin{eqnarray} \label{12}
\fl \nu(k)=\mu(k)-\frac{\pi}{2},\ooo \mu(k)=\pm \arctan\sqrt{\frac{R(k)}{T(k)}};
\end{eqnarray}
the corresponding amplitudes are
\begin{eqnarray*}
\fl A^{tr}_{in}=\sqrt{T}(\sqrt{T}\pm i\sqrt{R})=\frac{k(k\pm i\kappa)}{k^2+\kappa^2},\ooo
A^{ref}_{in}=\sqrt{R}(\sqrt{R}\mp i\sqrt{T})=\frac{\kappa(\kappa\mp i k)}{k^2+\kappa^2}.
\end{eqnarray*}
It is seen that $A^{tr}_{in}=A_{tr}$ and $A^{ref}_{in}=-A_{ref}$, for the lower sign; while $A^{tr}_{in}=A_{tr}^*$ and
$A^{ref}_{in}=-A_{ref}^*$, for the upper sign. For both roots $A^{tr}_{in}+A^{ref}_{in}=1$ and
$|A^{tr}_{in}|^2+|A^{ref}_{in}|^2=1$.

Considering only amplitudes corresponding to the lower sign, it is easy to show that the function $\Psi_{tot}^{L}(x,k)$
can be uniquely written as a superposition of the states $\Psi_{tr}^{L}(x,k)$ and $\Psi_{ref}^{L}(x,k)$:
\begin{eqnarray} \label{13}
\fl \Psi_{tot}^{L}(x,k)=A^{tr}_{in}(k)\Psi_{tr}^{L}(x,k)+A^{ref}_{in}(k)\Psi_{ref}^{L}(x,k)
\end{eqnarray}
(the amplitudes $A^{tr}_{in}$ and $A^{ref}_{in}$ which correspond to the upper sign in Exp. (\ref{12}) appear in the
expressions complex conjugate to Exps. (\ref{13})). Similarly, for the right incidence
\begin{eqnarray*}
\fl \Psi_{tot}^{R}(x,k)=A^{tr}_{in}(k)\Psi_{tr}^{R}(x,k)+A^{ref}_{in}(k)\Psi_{ref}^{R}(x,k).
\end{eqnarray*}

Thus, as it follows from (\ref{13}), the time-dependent scattering states $\Psi_{tot}^{L}$ and $\Psi_{tot}^{R}$ can be uniquely written as coherent superpositions
\begin{eqnarray} \label{14}
\fl \Psi_{tot}^L(x,t)=\Psi_{tr}^L(x,t)+\Psi_{ref}^L(x,t),\ooo \Psi_{tot}^R(x,t)=\Psi_{tr}^R(x,t)+\Psi_{ref}^R(x,t),
\end{eqnarray}
where
\begin{eqnarray} \label{16}
\fl \Psi_{tr}^{L,R}(x,t)=\frac{1}{\sqrt{2\pi}}\int_{-\infty}^\infty \mathcal{A}(k,t)A^{tr}_{in}(k)\Psi_{tr}^{L,R}(x,k)dk,\nonumber\\
\fl \Psi_{ref}^{L,R}(x,t)=\frac{1}{\sqrt{2\pi}}\int_{-\infty}^\infty
\mathcal{A}(k,t)A^{ref}_{in}(k)\Psi_{ref}^{L,R}(x,k)dk.
\end{eqnarray}
In particular, for the left incidence with the Gaussian function $\mathcal{A}_{in}(k)$
\begin{eqnarray} \label{316}
\fl \Psi_{tr}^{L}(x,t)=\Psi_{in}^L(x,t)-i\kappa G(x,t);\ppp \Psi_{ref}^{L}(x,t)=\left\{
\begin{array}{rl}
i\kappa [G(x,t)-G(-x,t)];\ooo x<0\\
\fl 0;\ooo x>0
\end{array} \right.
\end{eqnarray}

Let $T=\langle\Psi_{tr}^{L}|\Psi_{tr}^{L}\rangle$ and $R=\langle\Psi_{ref}^{L}|\Psi_{ref}^{L}\rangle$. Then
$\tilde{\Psi}_{tr}^{L}(x,t)=\Psi_{tr}^{L}(x,t)/\sqrt{T}$ and $\tilde{\Psi}_{ref}^{L}(x,t)=\Psi_{ref}^{L}(x,t)/\sqrt{R}$
are unite states that describe, respectively, the transmission and reflection subprocesses of the scattering process with
the left-sided incidence of a particle on the barrier.

\subsection{On the possibility of experimental investigation of subprocesses at all stages of scattering} \label{alt}

So, according to this model, the scattering state $\Psi_{tot}^{L}(x,t)$ with Gaussian in-asymptote $\Psi_{in}^{L}(x,t)$,
describing the scattering process with by left-hand fall of a particle on a barrier, can be written at all stages of
scattering as a superposition (\ref{14}) of two pure states $\Psi_{tr}^{L}(x,t)$ and $\Psi_{ref}^{L}(x,t)$ with norms
$\sqrt{T}$ and $\sqrt{R}$, respectively. This superposition has properties unusual for the Schrodinger formalism: it is a
mixed state, although the states $\Psi_{tr}^{L}(x,t)$ and $\Psi_{ref}^{L}(x,t)$ separated only in the limit $t\to\infty$.
At the previous stages, in the region $x<0$, they overlap and interfere with each other.

All this means that an experimental study of the transmission and reflection subprocesses in this area is possible only
with the help of {\it indirect} measurements. Obviously, this also applies to the determination of their characteristic
times (the impossibility of directly measuring the characteristic times is also emphasized in the work \cite{Ste0,Ste},
where the problem of indirect measurement is based on the formalism of conditional probabilities). And since the dynamics
of both subprocesses is restored from the in-asymptote $\Psi_{in}^L(x,t)$ and the out-asymptote $\Psi_{out}^{L}(x,t)$ and
$\Psi_{out}^{R}(x,t)$, wave functions $\Psi_{tr}^{L}(x,t)$ and $\Psi_{ref}^{L}(x,t)$ for $x<0$ should be written in terms
of these asymptotes, extending their dynamics to the whole time-axis.

Using Exps. (\ref{316}), we get
\begin{eqnarray} \label{31}
\fl \Psi_{tr}^{L}(x,t)=\Psi_{in}^L(x,t)+\Psi_{out}^{L}(-x,t), \ppp
\Psi_{ref}^{L}(x,t)=\Psi_{out}^L(x,t)-\Psi_{out}^{L}(-x,t).
\end{eqnarray}
Then, passing to the (normalized to unity) wave functions $\tilde{\Psi}_{tr}^{L}(x,t)=\Psi_{tr}^{L}(x,t)/\sqrt{T}$ and
$\tilde{\Psi}_{ref}^{L}(x,t)=\Psi_{ref}^{L}(x,t)/\sqrt{R}$, we calculate (in accordance with Section \ref{pure}) the
correspondinh probability fields $w_{tr}^L(x,t)$ and $w_{ref}^L(x,t)$, as well as the momentum fields $p_{tr}^L(x,t)$ and
$p_{ref}^L(x,t)$.

It is interesting to compare these functions with the corresponding fields $w_{tot}^L(x,t)$ and $p_{tot}^L(x,t)$, which
describe the whole process. Calculations are performed for the in-asymptote $\Psi_{in}^{L}(x,t)$ (see (\ref{53})) with
parameters corresponding to fig.~\ref{fig:fig1} for $\lambda = 0$. In this case $T\approx 0.2$ and $R\approx 0.8$. For the
very in-asymptote (\ref{53}))
\begin{eqnarray} \label{500}
\fl p_{in}^L(x,t)=\hbar \frac{2L^4k_0+(a+x)b t}{2(L^4+b^2 t^2)},\nonumber\\
\fl w_{in}^L(x,t)=\frac{c^2}{2\sqrt{L^4+b^2 t^2}} \exp\left(\frac{-L^2(x+a-2k_0 b t)^2}{2(L^4+b^2 t^2)}\right).
\end{eqnarray}
It easy to check that $\int_{-\infty}^\infty p_{in}^L(x,t) w_{in}^L(x,t) dx=\hbar k_0$.

\begin{figure}[h]
\begin{center}
\includegraphics[width=7.0cm,angle=0]{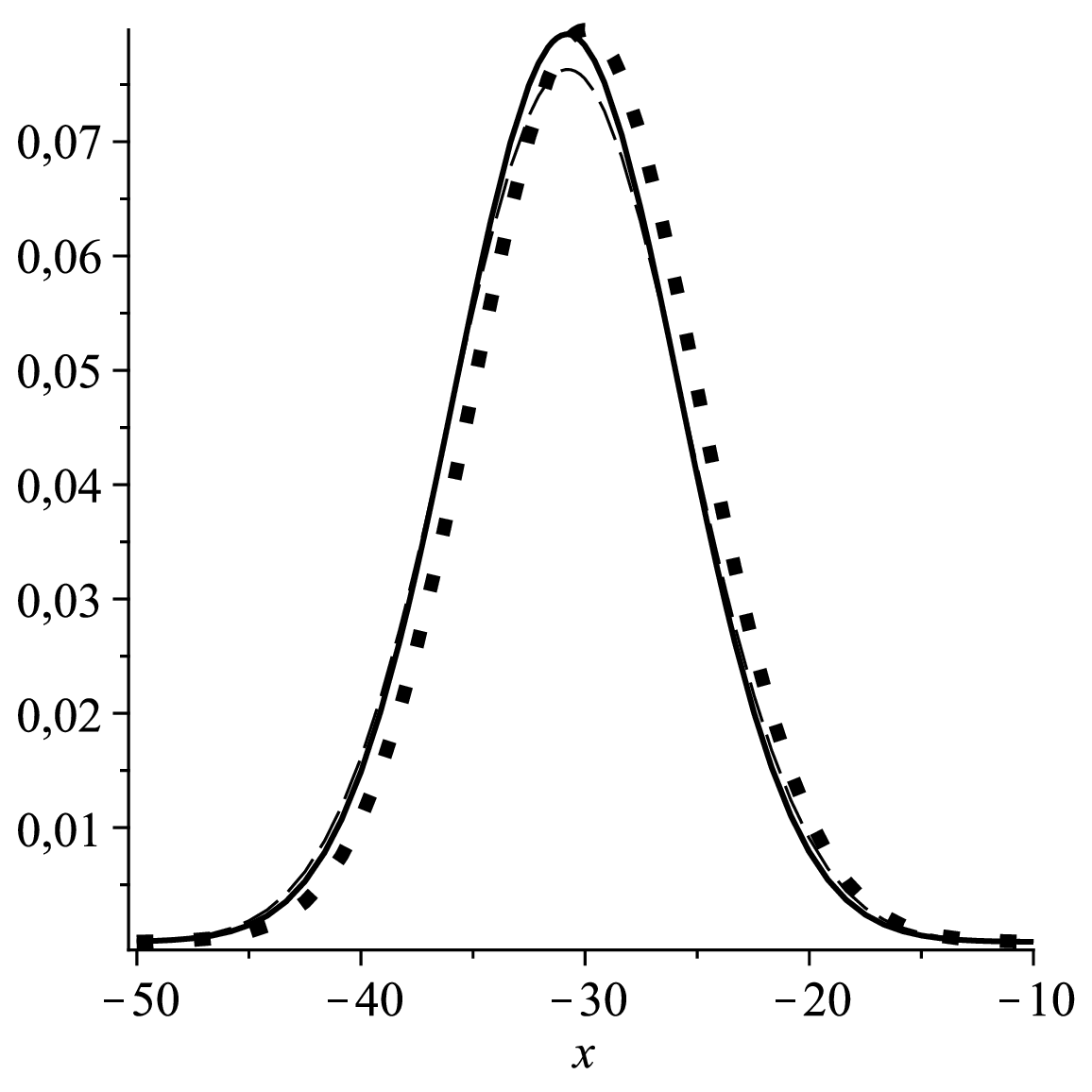}
\end{center}
\caption{$w_{tot}^L(x,0)$ (dotted line), $w_{tr}^L(x,0)$ (dashed line) and $w_{ref}^L(x,0)$ (solid line).}
\label{fig:fig2}
\end{figure}

\begin{figure}[h]
\begin{center}
\includegraphics[width=7.0cm,angle=0]{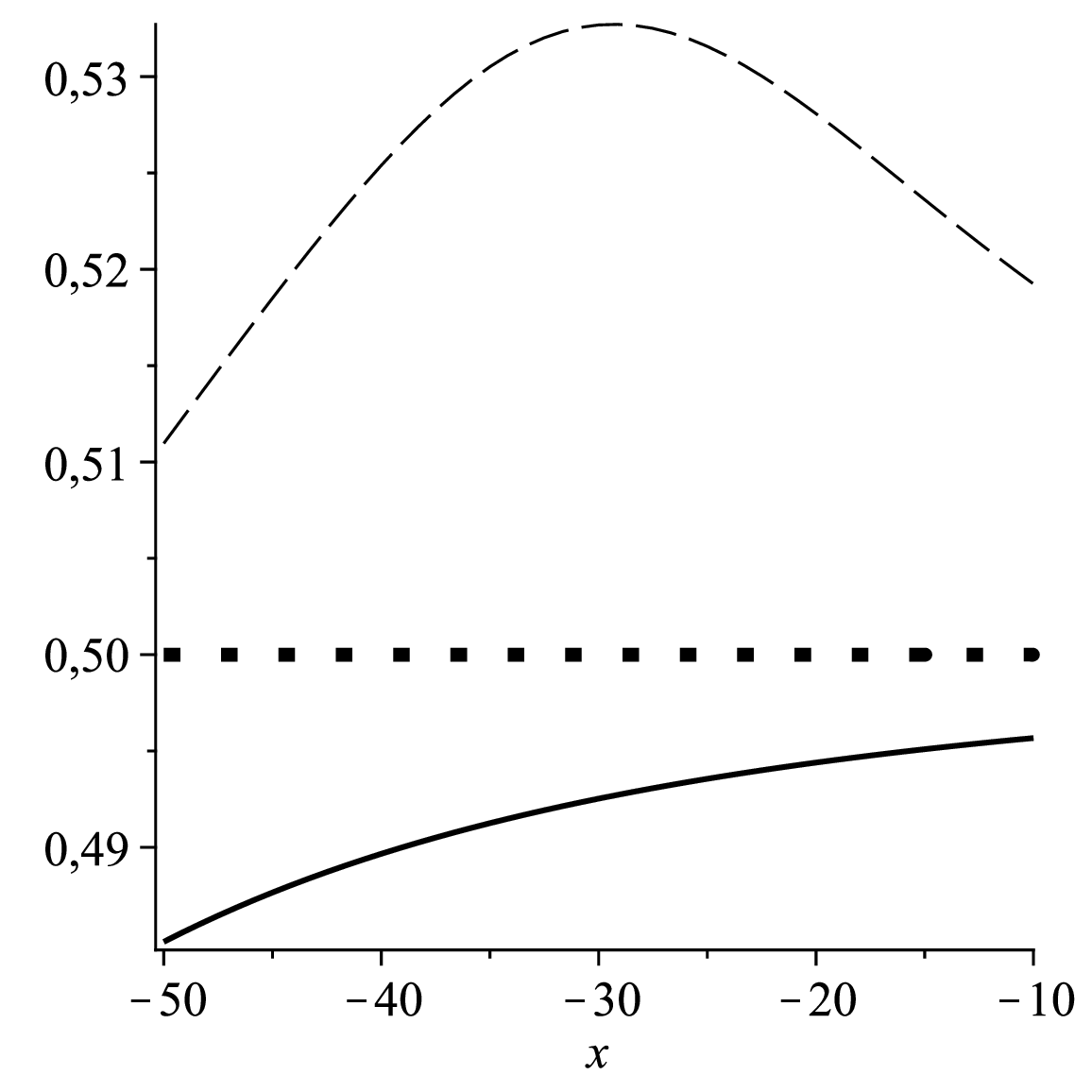}
\end{center}
\caption{$p_{tot}^L(x,0)$ (dotted line), $p_{tr}^L(x,0)$ (dashed line) and $p_{ref}^L(x,0)$ (solid line).}
\label{fig:fig3}
\end{figure}

Figs.~\ref{fig:fig2} and \ref{fig:fig3} show the calculation results for $t=0$.
As expected, at the initial moment of time all three probability fields (see Fig.~\ref{fig:fig2}) are noticeably different
from zero in the same spatial region. At all points of this region $p_{tr}^L(x,0)>p_{tot}^L(x,0)>p_{ref}^L(x,0)$, which
was also natural to expect.

\begin{figure}[h]
\begin{center}
\includegraphics[width=7.0cm,angle=0]{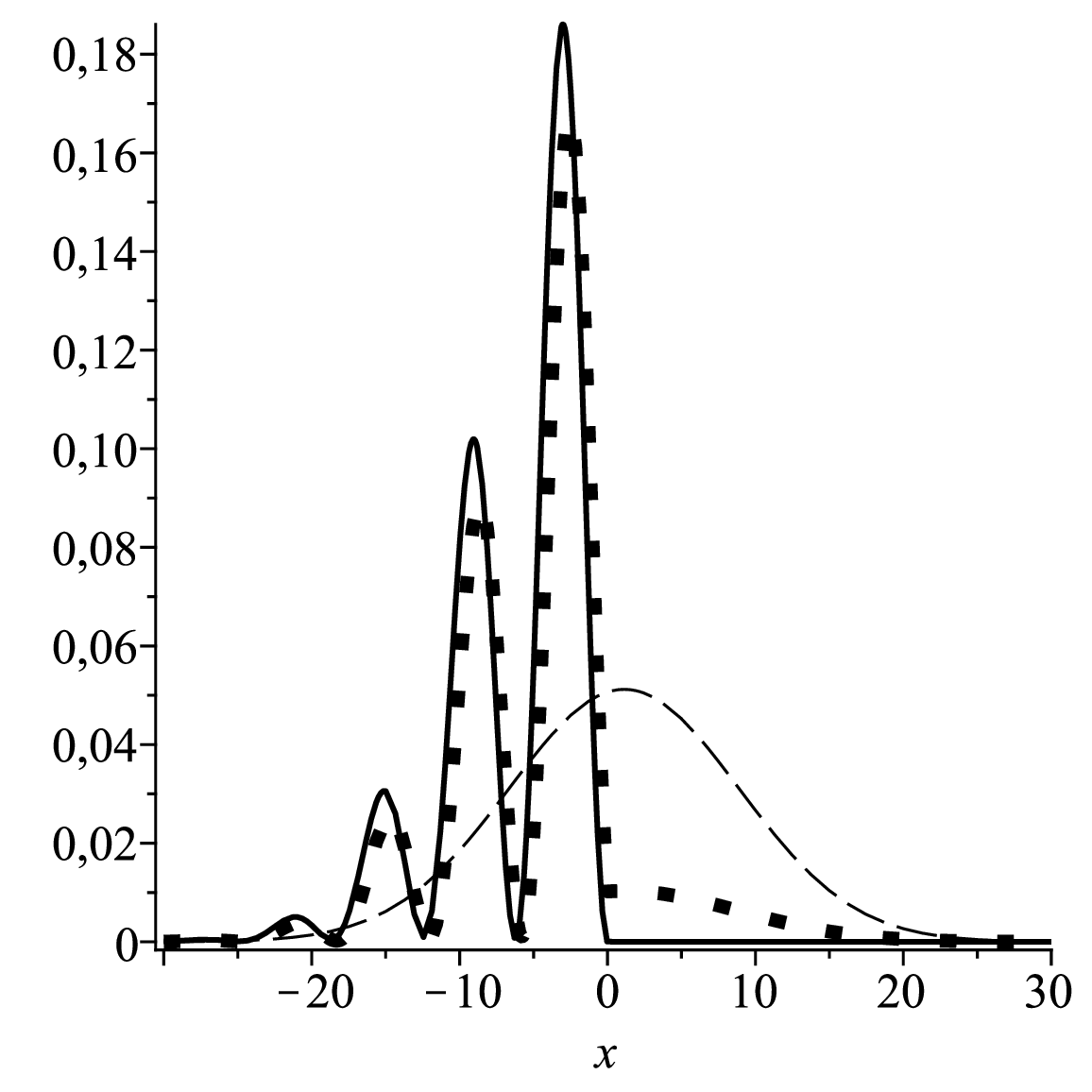}
\end{center}
\caption{$w_{tot}^L(x,60)$ (dotted line), $w_{tr}^L(x,60)$ (dashed line) and $w_{ref}^L(x,60)$ (solid line).}
\label{fig:fig4}
\end{figure}

\begin{figure}[h]
\begin{center}
\includegraphics[width=7.0cm,angle=0]{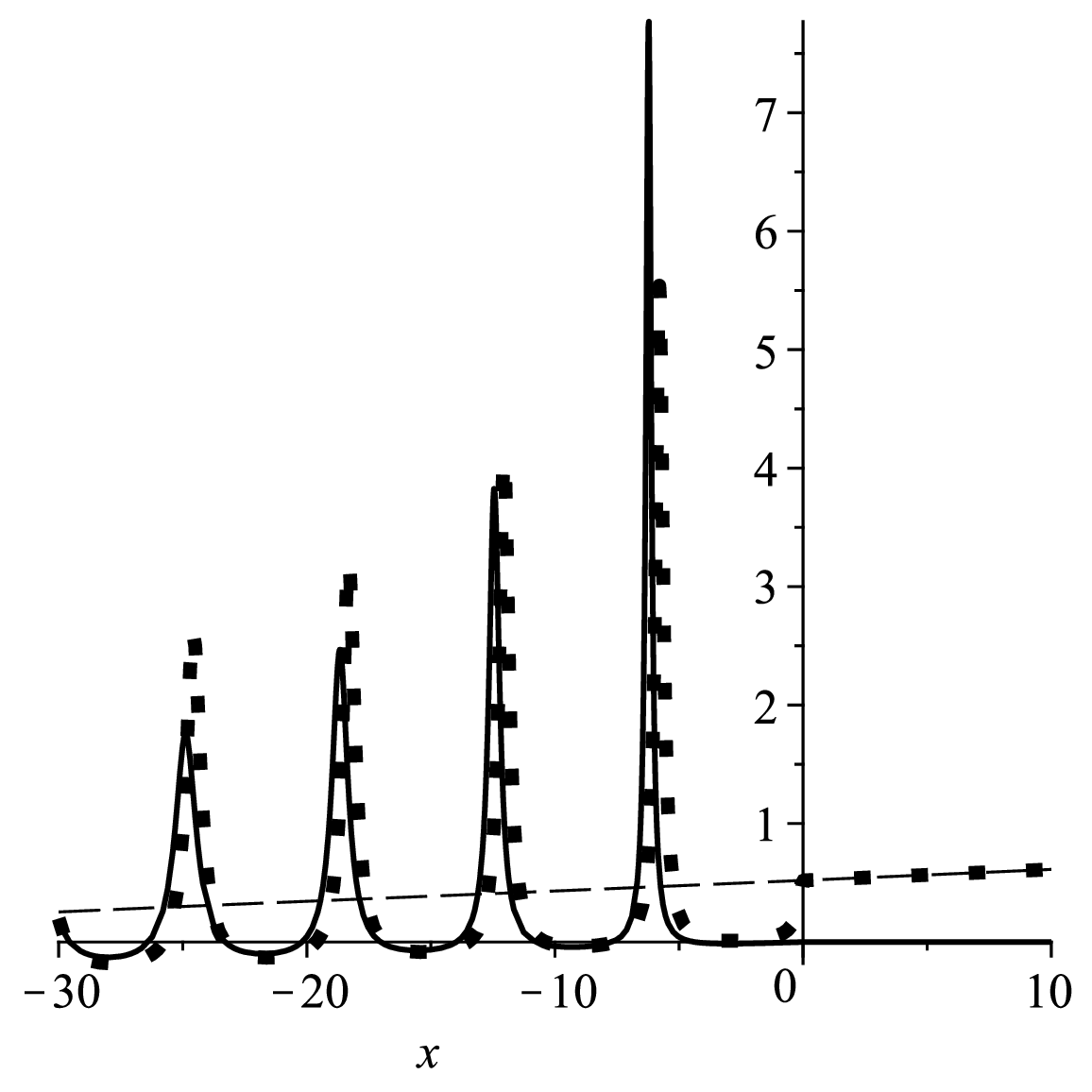}
\end{center}
\caption{$p_{tot}^L(x,60)$ (dotted line), $p_{tr}^L(x,60)$ (dashed line) and $p_{ref}^L(x,60)$ (solid line).}
\label{fig:fig5}
\end{figure}

The calculation results for $t=60$ are shown in figs.~\ref{fig:fig4} and \ref{fig:fig5}.

\section{Discussion and conclusions}

It is shown that the problem of scattering a particle by a one-dimensional $\delta$-potential is an analogue of the
thought experiment with two boxes, which Einstein used against the orthodox doctrine of the completeness of quantum
mechanics. In this regard, we presented our vision of this experiment. From our analysis it follows that the separation
principle, formulated by Einstein, is directed not so much against this doctrine (for this it is enough to consider the
experiment with one box), how much against the existing formulation of the principle of superposition, according to which
the state of a particle in boxes independent from each other should be considered as a pure quantum state. Moreover, this
formulation contradicts not only Einstein's separation principle, but also the definition of mixed states of quantum
mechanics itself. Thus, orthodox quantum mechanics is internally contradictory and requires a correction of the
superposition principle when describing quantum one-particle dynamics, during which ``Einstein boxes'' arise (i.e., when
states localized in spatial regions separated by an infinitely deep potential well). Apart from the scattering problem
considered, such a situation arises in the model \cite{Nun} of a one-dimensional hydrogen atom.

In the investigated scattering problem, ``Einstein's boxes'' arise when $t\to\infty$. In this limit, the scattering states
with asymptotically free dynamics lose the property of uniqueness and, therefore, develop in time in a non-unitary manner.
In other words, the formal Hamilton operator with $\delta$-potential is a non-self-adjoint operator, and the scattering
process with one-sided incidence of a particle on the barrier is a mixture of two subprocesses -- transmission and
reflection. An approach is presented for recovering the wave functions that describe these subprocesses, by the
in-asymptote of the entire process and the out-asymptotes describing the subprocesses. Based on these functions, it is
possible to indirectly measure the physical characteristics of each of their subprocesses for the first stages of
scattering.

Note that the question of the physical aspects of quantum dynamics in this scattering problem is solved, in our approach,
on the basis of new idea about the physical properties of a pure quantum ensemble, specified by the wave function.
According to modern quantum mechanics, the wave function imposes only statistical restrictions on the properties of the
ensemble. But this is far from the case. First, through the square of the modulus of the wave function (in the coordinate
representation), not only the distribution function of particles over coordinates (the 'probability field' of the
ensemble) is determined, but also its kinetic-energy field. Second, the phase of the wave function has a physical meaning
too -- it sets the momentum field of the ensemble. In this one-dimensional scattering problem, all these fields are
functions of two independent variables $x$ and $t$ (or, $p$ and $t$, in the momentum representation).

In other words, quantum mechanics not only does not prohibit the simultaneous measurement of the coordinate and momentum
of a particle (as well as kinetic energy), but also predicts the value of the momentum at that spatial point where the
particle will be (accidentally) detected. And this in no way contradicts the Heisenberg uncertainty principle, which
imposes restrictions on the standard deviations of the coordinates of particle momenta in a quantum ensemble, and not on
the measurement errors of these quantities.

The fact that the wave function predicts the fields of physical quantities that characterize a quantum ensemble suggests
that its physical properties are uniquely determined by external physical conditions (macroscopic physical context) under
which each member of the ensemble moves. We have also to note that the sets of fields in the $x$- and $p$-representations
are different: in the first case we have the $w(x,t)$-, $p(x,t)$- and $K(x,t)$-fields and so on; in the second case we
have the $w(p,t)$-, $x(p,t)$- and $x^2(p,t)$-fields and so on.

\end{document}